\documentclass[10pt,aps,prl,showpacs,superscriptaddress,twocolumn,citeautoscript]{revtex4-1}

\usepackage{graphicx}  
\usepackage{dcolumn}   
\usepackage{bm}        
\usepackage{amssymb}  
\usepackage{amsmath}
\usepackage{hyperref,color,braket}

\newcommand{\beq}{\begin{equation}}
\newcommand{\eeq}{\end{equation}}

\newcommand{\eps}{\varepsilon}
\newcommand{\subeqs}[1]{\begin{align} #1 \end{align}}
\begin{document}
\title{Zak's Phase in Non-Symmetric One-Dimensional Crystals}
\author{Marc Mart\'i-Sabat\'e}
\author{Dani Torrent}
\email{dtorrent@uji.es}
\affiliation{GROC, UJI, Institut de Noves Tecnologies de la Imatge (INIT), Universitat Jaume I, 12071, Castell\'o, (Spain)}
\date{\today}

\begin{abstract}
In this work, we derive some analytical properties of Berry's phase in one-dimensional quantum and classical crystals, also named Zak's phase, when computed with a Fourier basis. We show that Zak's phase can be divided in two terms: a global phase required to make the Bloch wave periodic in the Brillouin zone and an internal phase which measures the relative delay of the different Fourier terms within the Brillouin zone. While the former phase is dependent on the origin of coordinates of the unit cell, the latter is independent of it, so that it can be interpreted as an internal property of the band itself. We show that this internal phase is always zero for a symmetric crystal while it can take any value when this symmetry is broken, showing therefore that it can be interpreted as a measure of the assymetry of the band. Since for a symmetric crystal Zak's phase is entirely determined by the global part, we show that this can be easily calculated by means of the parity of the Fourier terms at the center and edge of the Brillouin zone, being therefore unnecessary the integration of the modes through the unit cell and the entire Brillouin zone. We provide numerical examples analyzing the internal part for both electronic and classical waves (acoustic or photonic). We analyze the weakest electronic potential capable of presenting asymmetry, as well as the double-Dirac delta potential, and in both examples it is found that the internal phase varies continuously as a function of a symmetry-control parameter, but it is zero when the crystal is symmetric. For classical waves, the layered material is analyzed. Although Zak's phase has been mainly studied in connection with the existence of edge states in finite crystals, we consider that the study of the internal phase can be more relevant to understand bulk properties of quantum and classical crystals.
\end{abstract}
\maketitle
The notion of Berry's phase\cite{berry1984quantal} has received increasing attention in condensed matter physics\cite{xiao2010berry}, since related quantities such as Berry connection and curvature are fundamental to understand the topological properties of matter\cite{hasan2010colloquium}. The richness of phenomena found for electrons in solids has also been exported to acoustic and photonic waves, and a wide variety of works have emerged in this realm\cite{lu2014topological,yang2015topological,ozawa2019topological}.

Berry's phase for energy bands in one dimensional crystals is also referred as Zak's phase, since in his seminal work\cite{zak1989berry} J. Zak first considered this important parameter. In his work, Zak showed that Berry's phase can take any value for a non-centrosymmetric crystal, while it can take only the values 0 and $\pi$ for a centrosymmetric one. This quantization of Zak's has been widely used in recent works \cite{xiao2010berry, delplace2011zak, atala2013direct,xiao2014surface,xiao2015geometric,choi2016simultaneous,esmann2018topological,yin2018band,ma2019topological} since its connection with the existence of interface states in finite crystals provides a powerful tool for the prediction of these states, however the non-symmetric crystal, where Zak's phase can in principle take any value, has received scarce attention. 

In this work, we use the plane wave expansion method (PWE) to derive some general properties of Zak's phase. We discuss several aspects of Zak's quantization condition for centrosymmetric crystals and then several examples are shown for non-centrosymmetric crystals. We show that Zak's phase defined in this way provides a unique quantifier of the chirality of a band, which is an internal parameter independent of the coordinates of the unit cell and can be calculated for each band independently. The results are easily exported to classical waves and numerical examples are also provided. 

Let us consider the one-dimensional Schrodinger  equation in normalized  units for a potential $v(x)$ and energy $\eps$,
\beq
-\psi''+v(x)\psi=\eps\psi,
\eeq
we will assume hereafter that the potential $v(x)$ is a periodic function of $x$ with period $a$, thus it can be expanded as
\beq
v(x)=\sum_m v_m e^{i2m\pi x/a}.
\eeq
Bloch's theorem allows us to express the wavefunction $\psi(x,k)$ as\cite{kittel1996introduction}
\beq
\label{eq:BW}
\psi(x,k)=e^{ikx}u(x,k)=e^{ikx}\sum_mu_m(k)e^{i2m\pi x/a},
\eeq
with $u(x,k)$ being a periodic function with the same period as $v(x)$, i.e., of period $a$, and $k$ is Bloch's wavenumber. Inserting the above expression into Schrodinger equation results in
\beq
\label{eq:MuE}
\sum_{m'}(k_m^2\delta_{mm'}+v_{m-m'})u_{m'}=\eps u_m,
\eeq
where we have defined $k_m=k+2\pi m/a$ and the explicit dependence of $u_m$ with $k$ has been ommited. From the above equation the eigenvalues $\eps=\eps(k)$ are obtained, what is known as the band structure. Since the potential $v(x)$ is real, its Fourier transform satisfies $v_{m-m'}=v_{m'-m}^*$, consequently the above eigenvalue equation is Hermitian and its eigenvalues are real. However, in the case of having a potential symmetric in the unit cell, that is, if we are able to find a unit cell such that $v(-x)=v(x)$, the properties of Fourier transform imply that $v_m$ is real as well, that is to say,
\beq
v(x)=v(-x)\rightarrow v_{m-m'}=v_{m'-m},
\eeq
and the eigenvalue equation \eqref{eq:MuE} is defined by means of a real and symmetric matrix, which has real eigenvalues and, most importantly, real eigenvectors. Naturally, we can obtain complex eigenvectors as well, since these can always be multiplied by an arbitrary constant, but this might add only a global trivial phase. These considerations are important for the calculation of the so-named Zak's phase\cite{zak1989berry}, which is Berry's phase for electrons in a periodic potential, and it is defined as the integral of the Berry connection $A(k)$,
\beq
\theta_0=\int_{-\pi/a}^{\pi/a}A(k)dk,
\eeq
with
\beq
A(k)=\frac{i}{a}\int_0^a dx u(x,k)\partial_k u^*(x,k).
\eeq
If we use the Fourier expansion of $u(x,k)$ we get
\beq
\frac{1}{a}\int_0^a u(x,k)\partial_k u^*(x,k)dx =\sum_mu_m(k)\partial_ku_m^*(k),
\eeq
and we arrive to the following expression for Zak's phase
\beq
\label{eq:ZP}
\theta_0=i\sum_m\int_{-\pi/a}^{\pi/a}u_m(k)\partial_ku_m^*(k)dk.
\eeq
However, the condition for having a well-defined phase from the above expression is that, as a function of $k$, the eigenvectors $u_m$ be smooth and continuous functions and that they make the Bloch wave \eqref{eq:BW} a periodic function in $k$, 
\beq
\psi(x,k+2\pi/a)=\psi(x,k),
\eeq
which is equivalent to impose
\beq
u_m(k+2\pi/a)=u_{m+1}(k).
\eeq
From equation \eqref{eq:MuE} we know that any solution $\hat{u}_m$ will satisfy
\beq
\label{eq:phi0um}
\hat{u}_m(k+2\pi/a)=e^{i\phi_0}\hat{u}_{m+1}(k),
\eeq
with $\phi_0$ being some phase which will depend on our choice of $\hat{u}_m(k)$ and the potential $v(x)$. Therefore, once we have found a solution $\hat{u}_m(k)$ being a continuous function of $k$, we need to impose the periodicity in $k$ of $\psi(x,k)$ by means of the correction of the phase $\phi_0$. This correction can be made taking into account that given a solution $\hat{u}_m(k)$ we can build solutions of \eqref{eq:MuE} as
\beq
\psi(x,k)=e^{ikx}e^{-ikd_0}\hat{u}(x-d_0',k),
\eeq
so that the condition $\psi(x,k+2\pi/a)=\psi(x,k)$ will be satisfied as long as
\beq
\label{eq:phi0}
\phi_0=2\pi\frac{d_0-d_0'}{a},
\eeq
which establishes a relationship between the $u_m$ and $\hat{u}_m$ vectors 
\beq
u_m(k)=e^{-ikd_0}e^{-2im\pi/ad_0'}\hat{u}_m(k).
\eeq
It is interesting to note that the phase factor due to $d_0$ moves Zak's phase, but not the one due to $d_0'$, this can be seen by introducing the above expression into \eqref{eq:ZP}, resulting in
\beq
\label{eq:ZPbis}
\theta_0=\hat{\theta}_0-2\pi\frac{d_0}{a},
\eeq
where we have defined $\hat{\theta}_0$ as Zak's phase computed with the $\hat{u}_m$ coefficients in equation \eqref{eq:ZP}, that is,
\beq
\label{eq:hZP}
\hat{\theta}_0=i\sum_m\int_{-\pi/a}^{\pi/a}\hat{u}_m(k)\partial_k\hat{u}_m^*(k)dk.
\eeq

The above results can be summarized as follows: given a periodic potential $v(x)$, applying Bloch theorem and solving for the periodic part $\hat{u}(x,k)$, we can build this function so that it is continuous in $k$, performing a phase jump of $\phi_0$ when $k$ is increased by $2\pi/a$. Then, for any pair of $d_0$ and $d_0'$ satisfying \eqref{eq:phi0} we will satisfy the periodicity of the Bloch function and, since $\phi_0$ is a fixed quantity we can set Zak's phase to any value according to \eqref{eq:ZPbis}. This result is a generalization of the classical statement that Zak's phase depends on the origin of the unit cell. Actually, it should be more correct to say that it depends on the origin of the unit cell and the phase of the Bloch function. 

It is interesting to mention that the term $\hat{\theta}_0$ will always be zero for a symmetric potential. The reason is that, if a center of symmetry can be found, as mentioned before, the eigenvectors $\hat{u}_m$ can be selected to be real, consequently equation \eqref{eq:hZP} will be trivially equal to zero. It means that Zak's phase will be entirely determined by the phase $\phi_0$ which relates the shifted Fourier components at the two borders of the Brillouin zone. This phase has been shown to be only 0 or $\pi$ according to the different values of the Bloch wave at the center or the border of the Brillouin zone when the origin of the unit cell is selected as one of the symmetry centers of the crystal\cite{zak1989berry}, so that the use of equation \eqref{eq:ZP} is no longer required, since we can determine Zak's phase by simply analyzing the partity of the Bloch wave at $k=0$ and $k=\pi/a$. This is explained below.

For a symmetric potential, Zak's phase is entirely determined by the relation \eqref{eq:phi0um}, which can be set as
\beq
u_m(\pi/a)=e^{i\phi_0}u_{m+1}(-\pi/a).
\eeq
At the band edge $k=\pi/a$, it is easy to see that selecting real eigenvectors these satisfy
\beq
\label{eq:Ppi}
u_{m}(\pi/a)=P_{\pi/a} u_{-m-1}(\pi/a)
\eeq
where $P_{\pi/a}=\pm 1$ is the parity of the Bloch wave at $k=\pi/a$. Similarly, it can be shown that at $k=0$ we have
\beq
\label{eq:P0}
u_m(0)=P_0u_{-m}(0),
\eeq
where now $P_0$ is the parity of the mode at $k=0$. Since we are selecting real eigenvectors, to avoid discontinuities at $k=0$ we can choose the following gauge,
\beq
u_m(k)=P_0u_{-m}(-k),
\eeq
consequently
\beq
u_m(\pi/a)=P_{\pi/a} u_{-m-1}(\pi/a)=P_{\pi/a}P_0 u_{m+1}(-\pi/a),
\eeq
which allow us to identify
\beq
e^{i\phi_0}=P_{\pi/a}P_0,
\eeq
so that, with this gauge, Zak's phase for a symmetric crystal will be $0$ or $\pi$ when the parity of the Bloch mode at $k=0$ and $k=\pi/a$ be equal or different, respectively. The above expression provides a simpler method of calculus of Zak's phase for symmetric crystals, since it is not required to perform the scalar product of the spatial functions or the integration through the Brillouin zone, since once obtained the $u_m$ eigenvectors at $k=0$ and $k=\pi/a$, equations \eqref{eq:P0} and \eqref{eq:Ppi} defines $P_0$ and $P_{\pi/a}$, respectively. Since this phase and its relationship with the existence of interface states has been widely studied in the literature, we will focus now on the internal part $\hat{\theta}_0$.

An interesting property of $\hat{\theta}_0$ is that it is independent on the origin of the unit cell $d_0'$, since any shift of this quantity introduces a $k$-independent phase in the eigenvectors resulting in an invariance of $\hat{\theta}_0$. We can therefore consider this quantity as a band property, whose physical meaning can be understood as follows: In order to build a continuous function $\hat{u}_m(k)$ we will set the phase of the $m=0$ component equal to zero. We can express each $\hat{u}_m$ component as a complex vector with modulus $|\hat{u}_m|$ and phase $\Phi_m$, being both quantities $k$-dependent. In that case, expression \eqref{eq:hZP} takes the form
\beq
\hat{\theta}_0=\sum_m\int_{-\pi/a}^{\pi/a}|u_m|^2\partial_k\Phi_mdk,
\eeq
which shows that $\hat{\theta}_0$ is a weighted average of the variation of the phase of each Fourier component with respect to the average field through the Brillouin zone. Obviously this choice is not unique, and we can similarly set the origin of the phase to the $\ell$-th component. This is equivalent to multiplication of each $\hat{u}_m$ computed previously by $e^{-i\Phi_\ell(k)}$, so that the new phase, labeled $\hat{\theta}_\ell$, will be
\beq
\label{eq:thetaell}
\hat{\theta}_\ell=\hat{\theta}_0-\int_{-\pi/a}^{\pi/a}\partial_k\Phi_\ell dk.
\eeq
Interestingly, any of these choices will result in a different $\phi_0$ but the same $\theta_0$. Therefore, while $\theta_0$ depends on the origin of coordinates of the unit cell but not on our choice for the calculation of the eigenvectors $u_m$, the phase $\hat{\theta}_0$ is independent of the origin of coordinates although it will depend on the selection of the reference Fourier term, although a clear relationship between the different choices is given by equation \eqref{eq:thetaell}. We provide below numerical examples of $\theta_0$, showing how it can characterize the asymetry of a band.

\begin{figure}[h!]
	\centering
	\includegraphics[width=\linewidth]{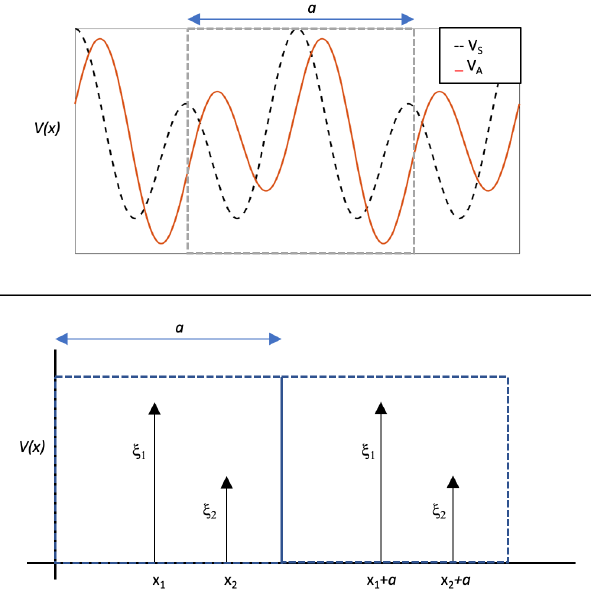}
	\caption{Electronic potentials analyzed in the numerical examples. Upper panel, weak asymmetric potential. Lower panel: Double-delta potential.}
	\label{fig:potentials}
\end{figure}

We have analyzed two different families of potentials, as shown in figure \ref{fig:potentials}. In the upper panel we can see the two examples of the ``weak potential'', which is defined as
\beq
v(x)=v_0+v_1\cos \frac{2\pi x}{a}+v_2\cos\left( \frac{4\pi x}{a}-\Phi_2\right).
\eeq
This corresponds to a potential where only the Fourier terms $m=0,\pm 1$ and $\pm2$ are different from zero, and it is the weakest potential that can be made non-symmetric, since setting $v_2=0$ always results in a symmetric potential. The parameter $\Phi_2$ controls the asymmetry of the potential, as can be seen in the upper panel where $v_S$ corresponds to $\Phi_2=0$ and $v_A$ corresponds to $\Phi_2=\pi/3$.

The lower panel of figure \ref{fig:potentials} shows the ``double delta potential'', which is defined as
\beq
v(x)=\xi_1\delta(x-d_1)+\xi_2\delta(x-d_2),
\eeq
with $\delta(x)$ being the Dirac delta function. The necessary condition for having a non-symmetric potential is that $\xi_1\neq\xi_2$, however this is not sufficient, since if for instance $d_1=0$ and $d_2=a/2$ we will have a symmetric crystal for all $\xi_1$ and $\xi_2$. 

\begin{figure}[h!]
	\centering
	\includegraphics[width=\linewidth]{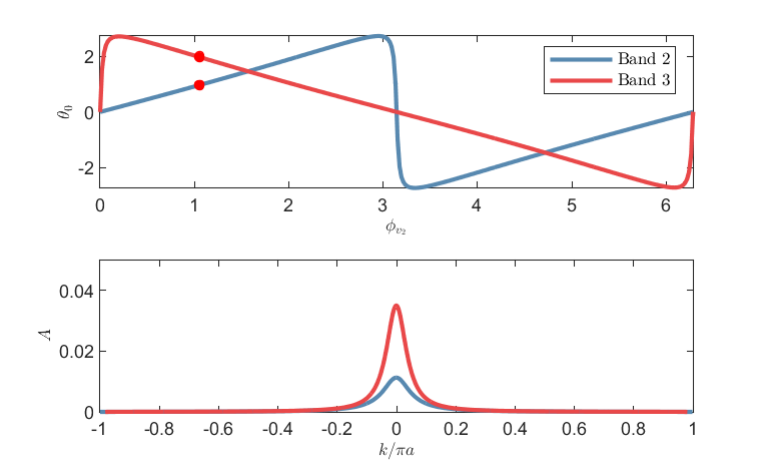}
	\caption{Upper panel: Internal part of Zak's phase for bands 2 and 3 for a weak non-symmetric potential as a function of phase $\Phi_2$. Lower panel: Berry connection for the configuration shown as red points in the upper panel.}
	\label{fig:ZW}
\end{figure}

Figure \ref{fig:ZW} upper panel shows $\hat{\theta}_0$ for the weak potential when $v_0=0, v_1=-1$ and $v_2=-2$. Results are shown as a function of $\Phi_2$ and for bands 2 and 3, since we found $\hat{\theta}_0=0$ for the lowest band. As expected, when $\Phi_2=0,\pi$ and $2\pi$ symmetry is recovered and $\hat{\theta}_0=0$, but in the full range it takes a regular behaviour. Lower panel shows Berry connection (computed with $\hat{u}_m$ as a function of $k/\pi a$ for the full Brillouin zone and for $\Phi_2=\pi/3$, marked as red dots in the upper panel. The main contribution $\hat{\theta}_0$ is due to small $k$.

\begin{figure}[h!]
	\centering
	\includegraphics[width=\linewidth]{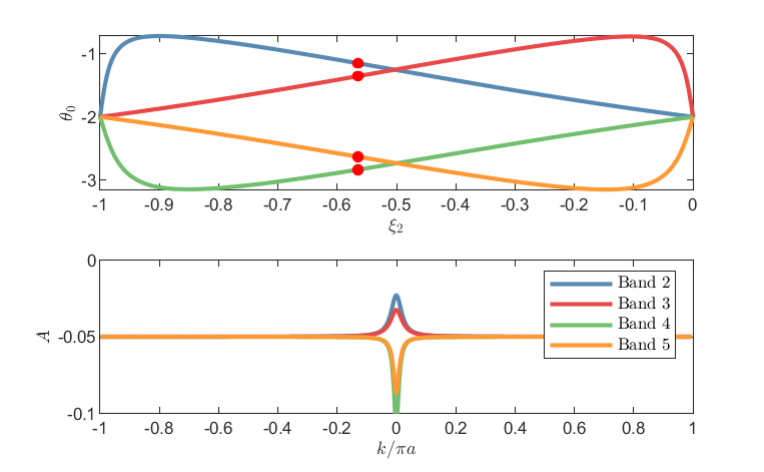}
	\caption{Upper panel: Internal part of Zak's phase for bands 2 to 5 for the double-delta potential as a function of $\xi_2$. Lower panel: Berry connection for the configuration shown as red points in the upper panel.}
	\label{fig:ZD}
\end{figure}

Figure \ref{fig:ZD},  upper panel shows $\hat{\theta}_0$ for the double-delta potential, as a function $\xi_2$ setting $\xi_1=-1$ and $d_2=-d_1=a/3$. Results are shown for bands 2 to 5 since again band 1 is trivial. We can see again how for the symmetric configurations $\xi_2=\xi_1$ and $\xi_2=0$ we recover the cancelation of $\hat{\theta}_0$, while it is a continuous function in all the other cases. As before, the lower panel shows Berry connection for $\xi_2=-0.56$ and the main contribution of the integral is around $k=0$. 

All the above considerations are also valid for acoustic or electromagnetic waves, where the wave equation takes the form
\beq
(\alpha(x)\psi')'=\beta(x)\omega^2\psi,
\eeq
where the $\alpha$ and $\beta$ coefficients are now the periodic functions of $x$. The eigenvalue equation is a generalized eigenvalue problem of the form
\beq
\sum_{m'}M_{mm'}u_{m'}=\omega^2\sum_{m'}N_{mm'}u_{m'},
\eeq
where the matrices $M$ and $N$ are
\subeqs{
M_{mm'}&=(k+\frac{2\pi m}{a})\alpha_{m-m'}(k+\frac{2\pi m'}{a}),\\
N_{nn'}&=\beta_{m-m'},
}
and the Berry phase is 
\beq
\label{eq:ZPB}
\theta_0=i\int_{-\pi/a}^{\pi/a}dk\frac{1}{a}\int_0^au(x)\beta(x)\partial_k u^*(x) dx ,
\eeq
which in terms of the $u_m$ coefficients is
\beq
\theta_0=i\sum_{m,m'}\int_{-\pi/a}^{\pi/a}u_m\beta_{m-m'}\partial_ku_{m'}^*dk.
\eeq

\begin{figure}[h!]
	\centering
	\includegraphics[width=\linewidth]{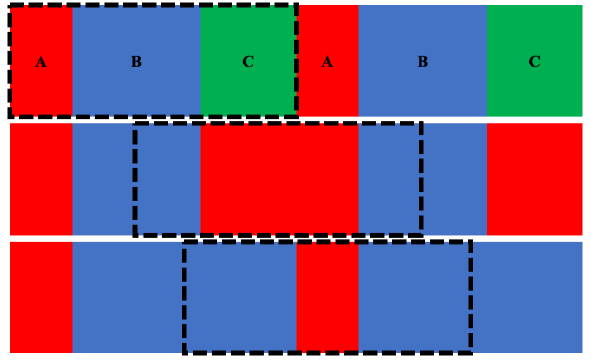}
	\caption{Unit cell for a three materials layered crystal (upper panel). It is shown that if $C=A$(mid panel) or $C=B$(lower panel) we can shift the unit cell to find a symmetric configuration.}
	\label{fig:ABC}
\end{figure}

The most commonly periodic material used for both acoustics and photonics is the layered material, where the unit cell is made of regions of given thickness with constant $\alpha$ and $\beta$ parameters. Figure \ref{fig:ABC} shows a three layers unit cell. Upper panel shows a clearly asymmetric configuration, but if we set material $C$ equal to $A$ (mid panel) or equal to $B$(lower panel), we can always find a symmetric unit cell and, as discussed before, we will have $\hat{\theta}_0=0$.

\begin{figure}[h!]
	\centering
	\includegraphics[width=\linewidth]{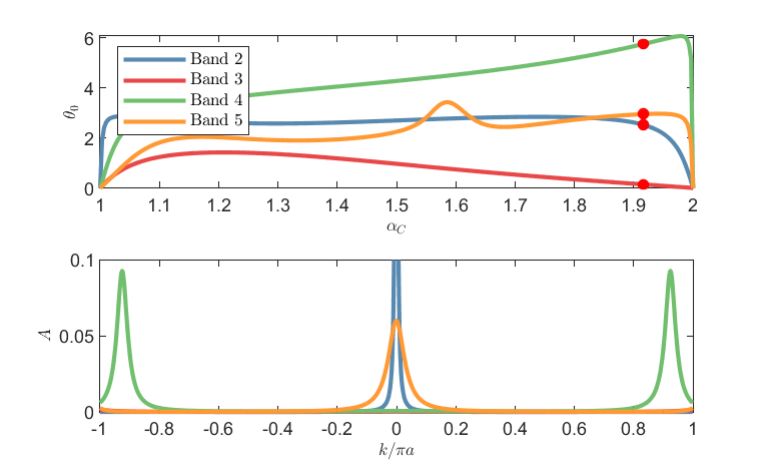}
	\caption{Upper panel: Internal part of Zak's phase for bands 2 to 5 for the triply-layered material as a function of $\alpha_C$. Lower panel: Berry connection for the configuration shown in red points in the upper panel.}
	\label{fig:ZABC}
\end{figure}

Finally, numerical examples for the three-layers material are shown if figure \ref{fig:ZABC}. We have selected $\beta(x)=1$, so that we modulate only $\alpha(x)$. We have set $\alpha_A=2,\alpha_B=1$ and we have plot Zak's phase as a function of $\alpha_C$. Each layer has a thickness $d_A=a/4,d_B=a/2,d_C=a/4$, respectively. Sweeping $\alpha_C$ from $\alpha_B$  to $\alpha_A$, we see how $\hat{\theta}_0$ (upper panel) takes finite values in all the range except at the initial and final points, which correspond to symmetric configurations. The lower panel shows Berry connection for $\alpha_C=1.92$, and it can be seen that for band 4 the contribution to $\hat{\theta}_0$ of Berry connection is due to its value near the border of the Brillouin zone.

In summary, we have shown that Zak's phase, when computed in a Fourier basis, can be divided in a global contribution plus an internal phase, which is in general independent of the origin of the unit cell. We have shown that if the crystal has a center of symmetry, the internal phase is trivially equal to zero, so that Zak's phase is entirely determined by the global part, and it is directly related with the change of parity of the Bloch wavefunction at the center and edge of the Brillouin zone. We have shown that the internal part of Zak's phase has a clear physical meaning as it measures the delay through the Brillouin zone of the different Fourier components of the wavefunction. The independence of the internal part of Zak's phase with the origin of coordinates of the unit cell shows that it might be an alternative mechanism for the characterization of bands and could provide additional information of the bulk properties of the crystal.

\section*{Acknowledgments}
D.T. acknowledges financial support through the ``Ram\'on y Cajal'' fellowship, under Grant No. RYC-2016-21188, and from the Ministry of Science, Innovation, and Universities, through Project No. RTI2018- 093921-AC42. M. M.-S. acknowledges financial support through the FPU program, under Grant No. FPU18/02725. Both authors acknowledge Andrew Norris and P. David Garc\'ia for useful and fruitful discussions. 


\begin{thebibliography}{17}
\expandafter\ifx\csname natexlab\endcsname\relax\def\natexlab#1{#1}\fi
\expandafter\ifx\csname bibnamefont\endcsname\relax
  \def\bibnamefont#1{#1}\fi
\expandafter\ifx\csname bibfnamefont\endcsname\relax
  \def\bibfnamefont#1{#1}\fi
\expandafter\ifx\csname citenamefont\endcsname\relax
  \def\citenamefont#1{#1}\fi
\expandafter\ifx\csname url\endcsname\relax
  \def\url#1{\texttt{#1}}\fi
\expandafter\ifx\csname urlprefix\endcsname\relax\def\urlprefix{URL }\fi
\providecommand{\bibinfo}[2]{#2}
\providecommand{\eprint}[2][]{\url{#2}}

\bibitem[{\citenamefont{Berry}(1984)}]{berry1984quantal}
\bibinfo{author}{\bibfnamefont{M.~V.} \bibnamefont{Berry}},
  \bibinfo{journal}{Proceedings of the Royal Society of London. A. Mathematical
  and Physical Sciences} \textbf{\bibinfo{volume}{392}}, \bibinfo{pages}{45}
  (\bibinfo{year}{1984}).

\bibitem[{\citenamefont{Xiao et~al.}(2010)\citenamefont{Xiao, Chang, and
  Niu}}]{xiao2010berry}
\bibinfo{author}{\bibfnamefont{D.}~\bibnamefont{Xiao}},
  \bibinfo{author}{\bibfnamefont{M.-C.} \bibnamefont{Chang}}, \bibnamefont{and}
  \bibinfo{author}{\bibfnamefont{Q.}~\bibnamefont{Niu}},
  \bibinfo{journal}{Reviews of modern physics} \textbf{\bibinfo{volume}{82}},
  \bibinfo{pages}{1959} (\bibinfo{year}{2010}).

\bibitem[{\citenamefont{Hasan and Kane}(2010)}]{hasan2010colloquium}
\bibinfo{author}{\bibfnamefont{M.~Z.} \bibnamefont{Hasan}} \bibnamefont{and}
  \bibinfo{author}{\bibfnamefont{C.~L.} \bibnamefont{Kane}},
  \bibinfo{journal}{Reviews of modern physics} \textbf{\bibinfo{volume}{82}},
  \bibinfo{pages}{3045} (\bibinfo{year}{2010}).

\bibitem[{\citenamefont{Lu et~al.}(2014)\citenamefont{Lu, Joannopoulos, and
  Solja{\v{c}}i{\'c}}}]{lu2014topological}
\bibinfo{author}{\bibfnamefont{L.}~\bibnamefont{Lu}},
  \bibinfo{author}{\bibfnamefont{J.~D.} \bibnamefont{Joannopoulos}},
  \bibnamefont{and}
  \bibinfo{author}{\bibfnamefont{M.}~\bibnamefont{Solja{\v{c}}i{\'c}}},
  \bibinfo{journal}{Nature photonics} \textbf{\bibinfo{volume}{8}},
  \bibinfo{pages}{821} (\bibinfo{year}{2014}).

\bibitem[{\citenamefont{Yang et~al.}(2015)\citenamefont{Yang, Gao, Shi, Lin,
  Gao, Chong, and Zhang}}]{yang2015topological}
\bibinfo{author}{\bibfnamefont{Z.}~\bibnamefont{Yang}},
  \bibinfo{author}{\bibfnamefont{F.}~\bibnamefont{Gao}},
  \bibinfo{author}{\bibfnamefont{X.}~\bibnamefont{Shi}},
  \bibinfo{author}{\bibfnamefont{X.}~\bibnamefont{Lin}},
  \bibinfo{author}{\bibfnamefont{Z.}~\bibnamefont{Gao}},
  \bibinfo{author}{\bibfnamefont{Y.}~\bibnamefont{Chong}}, \bibnamefont{and}
  \bibinfo{author}{\bibfnamefont{B.}~\bibnamefont{Zhang}},
  \bibinfo{journal}{Physical review letters} \textbf{\bibinfo{volume}{114}},
  \bibinfo{pages}{114301} (\bibinfo{year}{2015}).

\bibitem[{\citenamefont{Ozawa et~al.}(2019)\citenamefont{Ozawa, Price, Amo,
  Goldman, Hafezi, Lu, Rechtsman, Schuster, Simon, Zilberberg
  et~al.}}]{ozawa2019topological}
\bibinfo{author}{\bibfnamefont{T.}~\bibnamefont{Ozawa}},
  \bibinfo{author}{\bibfnamefont{H.~M.} \bibnamefont{Price}},
  \bibinfo{author}{\bibfnamefont{A.}~\bibnamefont{Amo}},
  \bibinfo{author}{\bibfnamefont{N.}~\bibnamefont{Goldman}},
  \bibinfo{author}{\bibfnamefont{M.}~\bibnamefont{Hafezi}},
  \bibinfo{author}{\bibfnamefont{L.}~\bibnamefont{Lu}},
  \bibinfo{author}{\bibfnamefont{M.~C.} \bibnamefont{Rechtsman}},
  \bibinfo{author}{\bibfnamefont{D.}~\bibnamefont{Schuster}},
  \bibinfo{author}{\bibfnamefont{J.}~\bibnamefont{Simon}},
  \bibinfo{author}{\bibfnamefont{O.}~\bibnamefont{Zilberberg}},
  \bibnamefont{et~al.}, \bibinfo{journal}{Reviews of Modern Physics}
  \textbf{\bibinfo{volume}{91}}, \bibinfo{pages}{015006}
  (\bibinfo{year}{2019}).

\bibitem[{\citenamefont{Zak}(1989)}]{zak1989berry}
\bibinfo{author}{\bibfnamefont{J.}~\bibnamefont{Zak}},
  \bibinfo{journal}{Physical review letters} \textbf{\bibinfo{volume}{62}},
  \bibinfo{pages}{2747} (\bibinfo{year}{1989}).

\bibitem[{\citenamefont{Delplace et~al.}(2011)\citenamefont{Delplace, Ullmo,
  and Montambaux}}]{delplace2011zak}
\bibinfo{author}{\bibfnamefont{P.}~\bibnamefont{Delplace}},
  \bibinfo{author}{\bibfnamefont{D.}~\bibnamefont{Ullmo}}, \bibnamefont{and}
  \bibinfo{author}{\bibfnamefont{G.}~\bibnamefont{Montambaux}},
  \bibinfo{journal}{Physical Review B} \textbf{\bibinfo{volume}{84}},
  \bibinfo{pages}{195452} (\bibinfo{year}{2011}).

\bibitem[{\citenamefont{Atala et~al.}(2013)\citenamefont{Atala, Aidelsburger,
  Barreiro, Abanin, Kitagawa, Demler, and Bloch}}]{atala2013direct}
\bibinfo{author}{\bibfnamefont{M.}~\bibnamefont{Atala}},
  \bibinfo{author}{\bibfnamefont{M.}~\bibnamefont{Aidelsburger}},
  \bibinfo{author}{\bibfnamefont{J.~T.} \bibnamefont{Barreiro}},
  \bibinfo{author}{\bibfnamefont{D.}~\bibnamefont{Abanin}},
  \bibinfo{author}{\bibfnamefont{T.}~\bibnamefont{Kitagawa}},
  \bibinfo{author}{\bibfnamefont{E.}~\bibnamefont{Demler}}, \bibnamefont{and}
  \bibinfo{author}{\bibfnamefont{I.}~\bibnamefont{Bloch}},
  \bibinfo{journal}{Nature Physics} \textbf{\bibinfo{volume}{9}},
  \bibinfo{pages}{795} (\bibinfo{year}{2013}).

\bibitem[{\citenamefont{Xiao et~al.}(2014)\citenamefont{Xiao, Zhang, and
  Chan}}]{xiao2014surface}
\bibinfo{author}{\bibfnamefont{M.}~\bibnamefont{Xiao}},
  \bibinfo{author}{\bibfnamefont{Z.}~\bibnamefont{Zhang}}, \bibnamefont{and}
  \bibinfo{author}{\bibfnamefont{C.~T.} \bibnamefont{Chan}},
  \bibinfo{journal}{Physical Review X} \textbf{\bibinfo{volume}{4}},
  \bibinfo{pages}{021017} (\bibinfo{year}{2014}).

\bibitem[{\citenamefont{Xiao et~al.}(2015)\citenamefont{Xiao, Ma, Yang, Sheng,
  Zhang, and Chan}}]{xiao2015geometric}
\bibinfo{author}{\bibfnamefont{M.}~\bibnamefont{Xiao}},
  \bibinfo{author}{\bibfnamefont{G.}~\bibnamefont{Ma}},
  \bibinfo{author}{\bibfnamefont{Z.}~\bibnamefont{Yang}},
  \bibinfo{author}{\bibfnamefont{P.}~\bibnamefont{Sheng}},
  \bibinfo{author}{\bibfnamefont{Z.}~\bibnamefont{Zhang}}, \bibnamefont{and}
  \bibinfo{author}{\bibfnamefont{C.~T.} \bibnamefont{Chan}},
  \bibinfo{journal}{Nature Physics} \textbf{\bibinfo{volume}{11}},
  \bibinfo{pages}{240} (\bibinfo{year}{2015}).

\bibitem[{\citenamefont{Choi et~al.}(2016)\citenamefont{Choi, Ling, Lee, Tsang,
  and Fung}}]{choi2016simultaneous}
\bibinfo{author}{\bibfnamefont{K.~H.} \bibnamefont{Choi}},
  \bibinfo{author}{\bibfnamefont{C.}~\bibnamefont{Ling}},
  \bibinfo{author}{\bibfnamefont{K.}~\bibnamefont{Lee}},
  \bibinfo{author}{\bibfnamefont{Y.~H.} \bibnamefont{Tsang}}, \bibnamefont{and}
  \bibinfo{author}{\bibfnamefont{K.~H.} \bibnamefont{Fung}},
  \bibinfo{journal}{Optics letters} \textbf{\bibinfo{volume}{41}},
  \bibinfo{pages}{1644} (\bibinfo{year}{2016}).

\bibitem[{\citenamefont{Esmann et~al.}(2018)\citenamefont{Esmann, Lamberti,
  Senellart, Favero, Krebs, Lanco, Carbonell, Lema{\^\i}tre, and
  Lanzillotti-Kimura}}]{esmann2018topological}
\bibinfo{author}{\bibfnamefont{M.}~\bibnamefont{Esmann}},
  \bibinfo{author}{\bibfnamefont{F.~R.} \bibnamefont{Lamberti}},
  \bibinfo{author}{\bibfnamefont{P.}~\bibnamefont{Senellart}},
  \bibinfo{author}{\bibfnamefont{I.}~\bibnamefont{Favero}},
  \bibinfo{author}{\bibfnamefont{O.}~\bibnamefont{Krebs}},
  \bibinfo{author}{\bibfnamefont{L.}~\bibnamefont{Lanco}},
  \bibinfo{author}{\bibfnamefont{C.~G.} \bibnamefont{Carbonell}},
  \bibinfo{author}{\bibfnamefont{A.}~\bibnamefont{Lema{\^\i}tre}},
  \bibnamefont{and} \bibinfo{author}{\bibfnamefont{N.~D.}
  \bibnamefont{Lanzillotti-Kimura}}, \bibinfo{journal}{Physical Review B}
  \textbf{\bibinfo{volume}{97}}, \bibinfo{pages}{155422}
  (\bibinfo{year}{2018}).

\bibitem[{\citenamefont{Yin et~al.}(2018)\citenamefont{Yin, Ruzzene, Wen, Yu,
  Cai, and Yue}}]{yin2018band}
\bibinfo{author}{\bibfnamefont{J.}~\bibnamefont{Yin}},
  \bibinfo{author}{\bibfnamefont{M.}~\bibnamefont{Ruzzene}},
  \bibinfo{author}{\bibfnamefont{J.}~\bibnamefont{Wen}},
  \bibinfo{author}{\bibfnamefont{D.}~\bibnamefont{Yu}},
  \bibinfo{author}{\bibfnamefont{L.}~\bibnamefont{Cai}}, \bibnamefont{and}
  \bibinfo{author}{\bibfnamefont{L.}~\bibnamefont{Yue}},
  \bibinfo{journal}{Scientific reports} \textbf{\bibinfo{volume}{8}},
  \bibinfo{pages}{1} (\bibinfo{year}{2018}).

\bibitem[{\citenamefont{Ma et~al.}(2019)\citenamefont{Ma, Xiao, and
  Chan}}]{ma2019topological}
\bibinfo{author}{\bibfnamefont{G.}~\bibnamefont{Ma}},
  \bibinfo{author}{\bibfnamefont{M.}~\bibnamefont{Xiao}}, \bibnamefont{and}
  \bibinfo{author}{\bibfnamefont{C.~T.} \bibnamefont{Chan}},
  \bibinfo{journal}{Nature Reviews Physics} \textbf{\bibinfo{volume}{1}},
  \bibinfo{pages}{281} (\bibinfo{year}{2019}).

\bibitem[{\citenamefont{Kittel et~al.}(1996)\citenamefont{Kittel, McEuen, and
  McEuen}}]{kittel1996introduction}
\bibinfo{author}{\bibfnamefont{C.}~\bibnamefont{Kittel}},
  \bibinfo{author}{\bibfnamefont{P.}~\bibnamefont{McEuen}}, \bibnamefont{and}
  \bibinfo{author}{\bibfnamefont{P.}~\bibnamefont{McEuen}},
  \emph{\bibinfo{title}{Introduction to solid state physics}},
  vol.~\bibinfo{volume}{8} (\bibinfo{publisher}{Wiley New York},
  \bibinfo{year}{1996}).

\bibitem[{\citenamefont{Kohn}(1959)}]{kohn1959analytic}
\bibinfo{author}{\bibfnamefont{W.}~\bibnamefont{Kohn}},
  \bibinfo{journal}{Physical Review} \textbf{\bibinfo{volume}{115}},
  \bibinfo{pages}{809} (\bibinfo{year}{1959}).

\end{thebibliography}
\end{document}